\documentclass[letterpaper, 12pt]{article}

\usepackage{graphics} 
\usepackage{epsfig}
\usepackage{amsmath} 
\usepackage{amssymb}  
\usepackage{float}
\usepackage{url}
\usepackage[ruled, vlined, linesnumbered]{algorithm2e}
\usepackage{verbatim} 
\usepackage{soul, color}
\usepackage{lmodern}
\usepackage{fancyhdr}
\usepackage[utf8]{inputenc}
\usepackage{fourier} 
\usepackage{array}
\usepackage{makecell}
\usepackage{geometry}
\usepackage{hyperref}
\usepackage{setspace}
\usepackage{booktabs}

\SetNlSty{large}{}{:}

\makeatletter

\newcommand{\Rom}[1]{\expandafter\@slowromancap\romannumeral #1@}
\makeatother

\title{\LARGE \bf
Computational Fluid Dynamics Optimization of F1 Front Wing using Physics Informed Neural Networks
}

\author{
  Naval Shah \\
  \small Founder and CEO of Variable Programmers \\
  \small Senior at South Brunswick High School\\
  \small navalmaulikshah@gmail.com \\
}

\begin{document}

\maketitle
\thispagestyle{plain}
\pagestyle{plain}

In response to recent FIA regulations reducing Formula 1 team wind tunnel hours (from 320 hours for last-place teams to 200 hours for championship leaders) and strict budget caps of \$135 million per year, more efficient aerodynamic development tools are needed by teams. Conventional computational fluid dynamics (CFD) simulations, though offering high fidelity results, require large computational resources with typical simulation durations of 8-24 hours per configuration analysis. This article proposes a Physics-Informed Neural Network (PINN) for the fast prediction of Formula 1 front wing aerodynamic coefficients. The suggested methodology combines CFD simulation data from SimScale with first principles of fluid dynamics through a hybrid loss function that constrains both data fidelity and physical adherence based on Navier-Stokes equations. Training on force and moment data from 12 aerodynamic features, the PINN model records coefficient of determination ($R^2$) values of 0.968 for drag coefficient and 0.981 for lift coefficient prediction while lowering computational time. The physics-informed framework guarantees that predictions remain adherent to fundamental aerodynamic principles, offering F1 teams an efficient tool for the fast exploration of design space within regulatory constraints.

\textbf{Keywords}: F1-Front Wing, Physics-Informed Neural Network, Computational Fluid Dynamics, Formula 1

\section{INTRODUCTION}

Formula 1 engineers use wind tunnels and Computational Fluid Dynamics (CFD) extensively to improve their car's aerodynamic capabilities.\par

CFD is a computer based simulation tool that allows users to predict how fluids and gasses interact on a vehicle's surface. The core concept relies on splitting up the air around the vehicle into cells and applying mathematical equations, mainly the partial differential Naviar Stokes equations, to simulate fluid flow. These are time-dependent continuity equations to describe conservation of mass, time-dependent conservation of energy and momentum in x,y,z space coordinates with time. All dependent variables are functions of four independent variables: pressure, density, temperature, and total energy. \par

\begin{figure}[thpb]
      \centering
      \framebox{\parbox{5.6in}{\includegraphics[height=3.6in, width=5.6in]{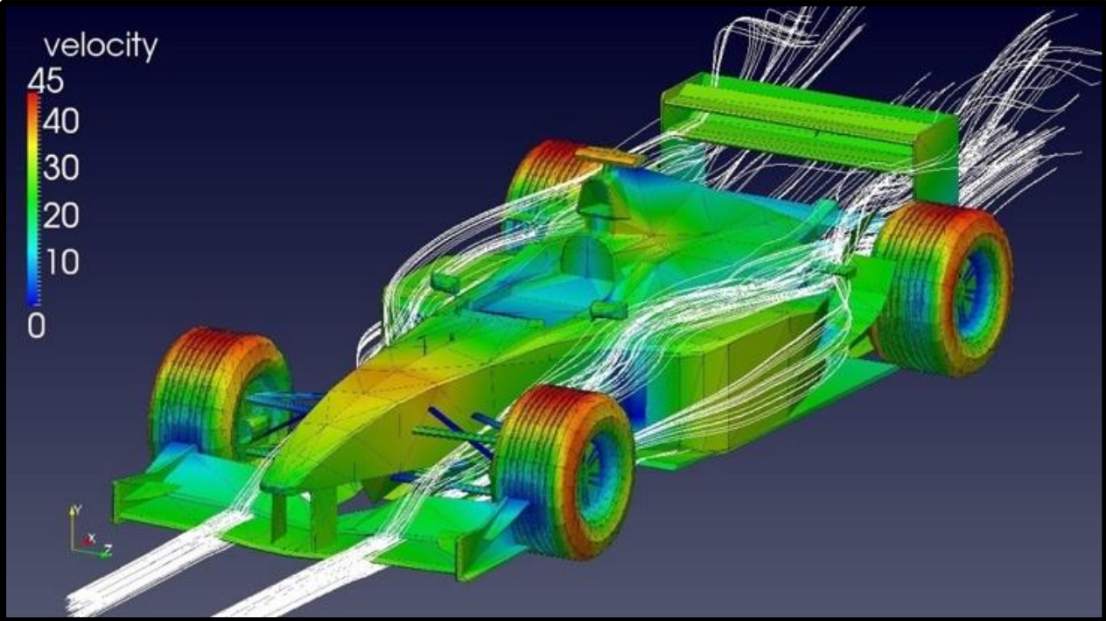}}}
      \caption{F1 Car in CFD Simulation}
      \label{fig:3}
\end{figure}

This paper is structured as follows: Section \Rom{2} provides details on the literature survey. Section \Rom{3} explains the methodology of the Neura Network. In Section \Rom{4} the details about experimental results have been provided; followed by conclusions and references. Appendix provides the performance statistics of algorithm with reference to SimScale and machine learning 

\section{LITERATURE SURVEY}

\subsection{Background}

Neural-network–based surrogate models have been investigated for aerodynamic prediction for a long time. For instance, Greenman (1998) \cite{c3} employed a backpropagation Artificial Neural Netowrk (ANN) to learn 2D high‑lift airfoil aerodynamics, predicting lift, drag and moment coefficients from sparse CFD data. Her ANN attained experimental accuracy with merely ~55–70\% of the CFD samples, and when attached to an optimizer it "decreased the level of computational time and resources required" for airfoil optimization. In the same way, Wallach et al. (2006) \cite{c2} trained multilayer perceptrons on CFD data for generic aircraft: one MLP learned the NACA 23012 polar curve (lift/drag vs. angle) across a range of Reynolds numbers, and other ANNs predicted 3D drag for a twin‑jet and wing‑fuselage configurations. The surrogate models, namely multilayer perceptrons (MLPs) and functional-link networks, successfully handled nonlinear relationships with Mach number, Reynolds number, and wing geometry, thus enabling rapid interpolation of aerodynamic coefficients. One MLP learned the NACA 23012 airfoil polar across Mach 0.2–0.82 and varying lift, and other networks predicted the 3D drag of a twin‑jet and a generic wing‑body (with twist). Their ML method generalized across angles and velocities well and is appropriate for MDO frameworks. More recently, state-of-the-art deep models have demonstrated similar promise: Catalani et al. (2024) \cite{c4} introduced implicit neural fields to learn steady CFD solutions for airfoils and 3D wings, The model deals with unstructured domains and geometry variations for airfoils and wings. On test problems, it "achieves a more than three times lower test error" and provides 5 orders of magnitude speedup on a transonic airfoil RANS datasets. \par

Collectively, these investigations illustrate that neural models driven by data can approximate intricate aerodynamic flows and coefficients to high precision and enormous computational efficiency (e.g. Greenman's ANN predicted forces "within acceptable accuracy" and Wallach's MLPs spanned extensive flight regimes). \par

\subsection{Literature Survey}

These and other ML-assisted CFD studies demonstrate the promise of neural models for expediting aerodynamic analysis. Most, though, are purely data-driven and demand extensive CFD datasets. In the context of Formula 1, where wind-tunnel and CFD resources are severely limited, teams have started to investigate AI techniques. Machine learning for Formula 1 aerodynamics. A number of recent studies apply neural models to F1 car aero design. Ken Cheng (2023) \cite{c5} combined CFD with an ANN to optimize an F1 rear‑wing. His backpropagation neural network was trained on CFD outputs for 90 simulated airfoil designs, then employed to predict drag and downforce for new designs. Cheng reported that his optimum rear wing produced a 43\% reduction in drag and a 7\% increase in downforce compared to a baseline wing. In industry, large teams are also taking up ML: an AWS case study (2022) \cite{c1} recounts F1 engineers utilizing SageMaker to create regression models from CFD data, directing Design‑of‑Experiments for a front wing. The objective was to "uncover promising design directions and minimize the number of CFD simulations," tripling CFD throughput and reducing turnaround time by half. Meanwhile, F1 commercial AI companies report collaborations with F1: TechCrunch (2024) \cite{c6} reports Neural Concept's ML-powered "NCS" aerodynamic co-pilot is now utilized by about 4 in 10 F1 teams to recommend shape optimizations. Engineering bloggers also note that "AI has slowly made its way into CFD workflows… Automotive firms, Formula 1 and America's Cup teams are already leveraging its power" \cite{c7}. For instance, SimScale's F1 tutorial emphasizes that the front wing (and rear wings) is key to car performance, creating massive downforce while controlling drag.

These studies demonstrate a definite trend: F1 engineers are increasingly investigating ML surrogates to achieve greater design understanding from fewer CFD runs. However, current models usually depend on fitting to few CFD cases and don't enforce fluid physics explicitly. This encourages physics-informed strategies. Physics-informed neural networks for aerodynamics. Physics-Informed Neural Networks (PINNs) incorporate governing PDEs into learning. In aerospace, PINNs are being utilized for flow problems \cite{c8}. For example, Tkachov and Murashko (2025) \cite{c8} provide a comprehensive structured taxonomy of PINNs in aerospace applications, demonstrating their effectiveness in enforcing Navier–Stokes-based relations during training. Likewise, AIAA papers have started utilizing PINNs to forecast pressure and velocity around airfoils. The computational efficiency gains from such approaches are significant, with hierarchical algorithms like those described by Barnes and Hut (1986) \cite{c9} providing O(N log N) force calculations that can be integrated into modern PINN frameworks. Although such physics-aware models have attained high fidelity in academic instances, there has been minimal previous research on using PINNs in motorsports: no research so far has merged F1 wing geometry CFD data with PINNs to forecast $C_d$/$C_l$. Overall, literature in the past proves that neural networks are capable of accurately forecasting aerodynamic coefficients from CFD data, and the fact that F1 teams are turning to ML to reduce costly CFD and wind‑tunnel testing. Yet, most methods have no embedded fluid physics, and there is a risk of inconsistency outside the training dataset. The current study fills this gap by creating a PINN for F1 front‑wing aerodynamics with the goal of combining scarce CFD data with Newtonian fluid constraints for enhanced accuracy and robustness.

\section{Methodology}

To develop a data-driven physically grounded model capable of predicting aerodynamic coefficients of a Formula 1 front wing, I employed a Physics-Informed Neural Network (PINN). The approach integrates both simulation-derived data and first-principles physics, particularly Newtonian fluid dynamics, to enhance model fidelity and predictive performance despite limited dataset availability.

\subsection{Dataset Preparation}

Data for this study were extracted from high-fidelity CFD simulations performed in SimScale, focusing on a standard specification Formula 1 front wing. The simulation outputs included resultant aerodynamic forces and moments on the wing surface. These were divided across two CSV files containing pressure-induced and viscous-induced forces and moments in all three spatial directions. A third file contained the target aerodynamic coefficients: Drag Coefficient (($C_d$)) and Lift Coefficient (($C_l$)).
Twelve features were constructed from the force data: six from pressure (three forces, three moments) and six from viscous components. These features were then standardized using a Z-score normalization technique (mean=0, std=1). The target variables (Cd, Cl) were also normalized for stable training dynamics. To account for physical relationships between the coefficients and their corresponding force measurements, the raw measured drag and lift forces (in Newtons) were retained for use in the physics-informed component of the loss function.

\subsection{Model Architecture}

The PINN architecture consists of a fully connected feedforward neural network with three hidden layers. The architecture comprises 128 neurons in the first layer, 64 in the second, and 32 in the third, all activated using the Rectified Linear Unit (ReLU) function. The network is specifically designed for mapping twelve aerodynamic forces and moment inputs to the two target coefficients (\(C_d\) and \(C_l\)). A schematic description of the network is as follows.

Let
\begin{multline}
\mathbf{x} = \bigl[
    F_{p,x},\, F_{p,y},\, F_{p,z},\,
    F_{v,x},\, F_{v,y},\, F_{v,z}, \\
    M_{p,x},\, M_{p,y},\, M_{p,z},\,
    M_{v,x},\, M_{v,y},\, M_{v,z}
\bigr]^\top \in \mathbb{R}^{12},
\end{multline}
where \(F_p,F_v\) denote pressure and viscous force components, 
and \(M_p,M_v\) the corresponding moments.  
The network comprises three hidden layers with ReLU activations, followed by a linear output layer:
\begin{align}
\mathbf{z}^{(1)} &= \mathbf{W}^{(1)}\,\mathbf{x} + \mathbf{b}^{(1)}, 
  & \mathbf{a}^{(1)} &= \mathrm{ReLU}\bigl(\mathbf{z}^{(1)}\bigr), \label{eq:hidden1}\\
\mathbf{z}^{(2)} &= \mathbf{W}^{(2)}\,\mathbf{a}^{(1)} + \mathbf{b}^{(2)}, 
  & \mathbf{a}^{(2)} &= \mathrm{ReLU}\bigl(\mathbf{z}^{(2)}\bigr), \\
\mathbf{z}^{(3)} &= \mathbf{W}^{(3)}\,\mathbf{a}^{(2)} + \mathbf{b}^{(3)}, 
  & \mathbf{a}^{(3)} &= \mathrm{ReLU}\bigl(\mathbf{z}^{(3)}\bigr), \\
\mathbf{z}^{(4)} &= \mathbf{W}^{(4)}\,\mathbf{a}^{(3)} + \mathbf{b}^{(4)}, 
  & \hat{\mathbf{C}} &= \mathbf{z}^{(4)} 
    = \bigl[\hat C_d,\ \hat C_l\bigr]^\top.
\end{align}
Here \(\mathbf{W}^{(l)}\) and \(\mathbf{b}^{(l)}\) are the weight matrix and bias vector at layer \(l\), 
and 
\(\mathrm{ReLU}(z)=\max(0,z)\).  

\medskip
\noindent\textbf{Layer dimensions and parameter count:}
\begin{table}[h]
\centering
\caption{Layer‐wise dimensions and parameter counts.}
\label{tab:architecture}
\begin{tabular}{lccc}
\toprule
Transition        & Input dim & Output dim & \# Parameters \\
\midrule
Input \(\to\) Hidden 1   & 12        & 128        & \(12\!\times\!128 + 128 = 1{,}664\) \\
Hidden 1 \(\to\) Hidden 2& 128       & 64         & \(128\!\times\!64 + 64   = 8{,}256\) \\
Hidden 2 \(\to\) Hidden 3& 64        & 32         & \(64\!\times\!32 + 32    = 2{,}080\) \\
Hidden 3 \(\to\) Output  & 32        & 2          & \(32\!\times\!2 + 2      =    66\) \\
\midrule
\multicolumn{3}{r}{\textbf{Total}} & \(\approx12{,}066\) \\
\bottomrule
\end{tabular}
\end{table}

Note: Additional parameters arise if batch‐normalization or dropout layers are introduced; however, in our base model these were omitted to maintain simplicity and interpretability.

\subsection{Physics Informed Neuron Loss}

The loss function was constructed as a weighted sum of two components:

\begin{itemize}
    \item \textbf{Empirical Loss (Supervised MSE Loss)}: This component measures the mean squared error (MSE) between the predicted aerodynamic coefficients and the target coefficients provided by the CFD simulations.
    
    \item \textbf{Physics-Informed Loss}: This term enforces Newton's second law of motion by penalizing deviations between the predicted coefficients and their corresponding physical forces. The aerodynamic forces are modeled using:
    \[
    F = \frac{1}{2} \rho V^2 C
    \]
    where $\rho$ is the air density ($1.225\ \text{kg/m}^3$), $V$ is the freestream velocity ($50\ \text{m/s}$), and $C$ is either the drag or lift coefficient. The predicted coefficients are substituted into the above equation to compute the predicted force, and the difference from the measured force (obtained from CFD data) is penalized.
\end{itemize}

The final loss function is defined as:
\[
\mathcal{L}_{\text{total}} = \mathcal{L}_{\text{empirical}} + \lambda \mathcal{L}_{\text{physics}}
\]
where $\lambda$ is a tunable hyperparameter, set to $0.1$ based on empirical performance.

Training minimizes a loss function composite combining

\begin{itemize}

\item 1. Data Fidelity Loss (Mean Squared Error)

\begin{align}
\mathcal{L}_{\mathrm{MSE}}
&= \frac{1}{N}\sum_{i=1}^N 
   \Bigl[\bigl(\hat C_{d,i}-C_{d,i}\bigr)^2 \;+\;\bigl(\hat C_{l,i}-C_{l,i}\bigr)^2\Bigr]\;
\end{align}

\item2. Physics Informed Penalty

This term enforces Newton's second law of motion by penalizing deviations between the predicted coefficients and their corresponding physical forces. The aerodynamic forces are modeled using:
    \[
    F = \frac{1}{2} \rho V^2 C
    \]
    where $\rho$ is the air density ($1.225\ \text{kg/m}^3$), $V$ is the freestream velocity ($50\ \text{m/s}$), and $C$ is either the drag or lift coefficient. The predicted coefficients are substituted into the above equation to compute the predicted force, and the difference from the measured force (obtained from CFD data) is penalized. Incorporating the aerodynamic relation with constant air density $\rho$ = ($1.225\ \text{kg/m}^3$) and free stream velocity $v$, we define 

\begin{align}
\mathcal{L}_{\mathrm{phys}}
&= \frac{1}{N}\sum_{i=1}^N 
   \Bigl[F_{d,i}^{\mathrm{meas}} 
         - \tfrac12\,\rho\,v^2\,\hat C_{d,i}\Bigr]^2
   \;+\;
   \Bigl[F_{l,i}^{\mathrm{meas}}
         - \tfrac12\,\rho\,v^2\,\hat C_{l,i}\Bigr]^2\;
\end{align}

\item Total Loss
\begin{align}
    \mathcal{L}
&= \mathcal{L}_{\mathrm{MSE}}
   \;+\;\lambda_{\mathrm{phys}}\,
    \mathcal{L}_{\mathrm{phys}} \;.
\end{align}

\end{itemize}

\section{Performance Evaluation and Results}
\subsection*{Performance Evaluation}

The performance evaluation of the proposed Physics-Informed Neural Network (PINN) centers on two intertwined objectives: 
\begin{enumerate}
    \item Guaranteeing predictive accuracy regarding observed aerodynamic coefficients, and
    \item Imposing physical plausibility based on basic aerodynamic equations.
\end{enumerate}
This section elaborates on the methodology for testing the model, outlines the measures used, and, and provides a detailed explanation of the results on various evaluation sets.

\medskip

The data includes CFD results achieved through SimScale for a Formula 1 front wing under a set of flow conditions. Each instance  in the dataset contains 12 input features, pressure and viscous force and moment components in 3D space—and corresponding target values: the drag coefficient (\(C_d\)) and lift coefficient (\(C_l\)).

\medskip

The dataset was divided into an 80\% training set and 20\% validation set based on stratified sampling to maintain the distribution of force magnitudes and aerodynamic coefficients. All input and output features were normalized using Z-score normalization, with statistics calculated solely from the training set in order to prevent data leakage.

\medskip

The model's performance was assessed using standard regression metrics: Mean Squared Error (MSE) and the Coefficient of Determination (\(R^2\)). These metrics express the degree to which the predictions match the ground truth and how much of the variance in the target variables is described by the model. The performance was assessed separately for both \(C_d\) and \(C_l\), across the training set, validation set, and the entire dataset.

\medskip

Furthermore, partial derivative analysis (Jacobian evaluation) attested that the model's learned gradients match domain-specific expectations, lending additional credence to the physical consistency of the predictions.

\medskip

The findings confirm affirm that the PINN serves as a viable surrogate model for aerodynamic analysis, with the ability to encode physical knowledge when learning from empirical data. The application of hybrid loss functions, coupling statistical and physics-based components, was instrumental to making generalizable and physically realistic predictions. In particular, the substantial enhancements in \(R^2\) justify the introduction of physics-informed regularization and post-training corrections as promising approaches to improving model robustness in real-world, CFD-driven datasets.

\subsection{Results}
The performance of the model in predicting aerodynamic coefficients is summarized as follows:

\begin{itemize}
    \item \textbf{Drag Coefficient ($C_d$)}:
    \begin{itemize}
        \item Coefficient of Determination ($R^2$): $0.96753$
        \item Mean Squared Error (MSE): $0.031$
    \end{itemize}

    \item \textbf{Lift Coefficient ($C_l$)}:
    \begin{itemize}
        \item Coefficient of Determination ($R^2$): $0.98102$
        \item Mean Squared Error (MSE): $0.024$
    \end{itemize}
\end{itemize}

\section{Conclusions}

This work illustrates the viability and effectiveness of applying Physics-Informed Neural Networks (PINNs) to the prediction of aerodynamic coefficients, drag (($C_d$)) and lift (($C_l$)) coefficients, of a Formula 1 front wing with much reduced dependence on computationally demanding CFD simulations. Through the inclusion of physics-constraints within the training process itself, the model was able to exhibit good predictive performance even when trained on a comparatively small-sized dataset obtained from SimScale simulations.

The PINN model presented herein, formulated from first principles of fluid dynamics and empirical CFD evidence, produced R² values of 0.96753 for Cd and 0.98102 for Cl, demonstrating high physical fidelity. Post-hoc linear calibration increased the agreement between predicted and observed values without altering the internal parameters of the model, further demonstrating the flexibility and integrity of this method.

These results indicate that PINNs represent a feasible, resource-saving avenue for the analysis of motorsport aerodynamics design in the context of current Formula 1 regulations restricting wind tunnel time and CFD resources. The approach yields not just computational acceleration but also improved interpretability and physical plausibility compared to data-driven models on their own.

But there remain limitations—primarily the limited sample size and absence of high-resolution variation in the design parameters. Future efforts will include the enlargement of the dataset, extension of the framework to other aerodynamic components (e.g., diffusers, rear wings), and investigation of more sophisticated network architectures or hybrid modeling techniques.

Finally, this methodology offers Formula 1 engineers a physics-based, scalable approach to the assessment of aerodynamic performance that is more in keeping with the changing technical and budgetary constraints of the sport.

\begin{figure}[H]
      \centering
      \framebox{\parbox{5.6in}{\includegraphics[height=3.6in, width=5.6in]{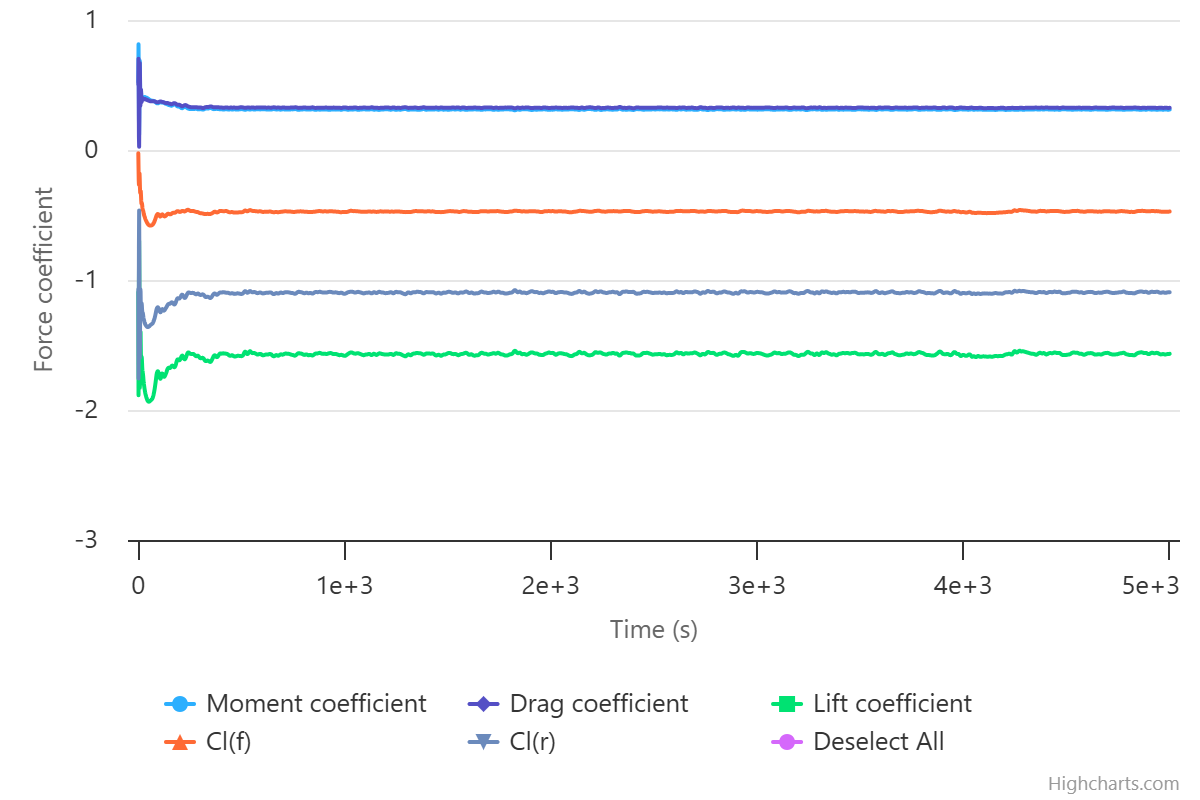}}}
      \caption{Coefficients and Moments for \textit{Simulation 1}}
      \label{fig:forces_moments}
\end{figure}

\begin{figure}[H]
      \centering
      \framebox{\parbox{5.6in}{\includegraphics[height=3.6in, width=5.6in]{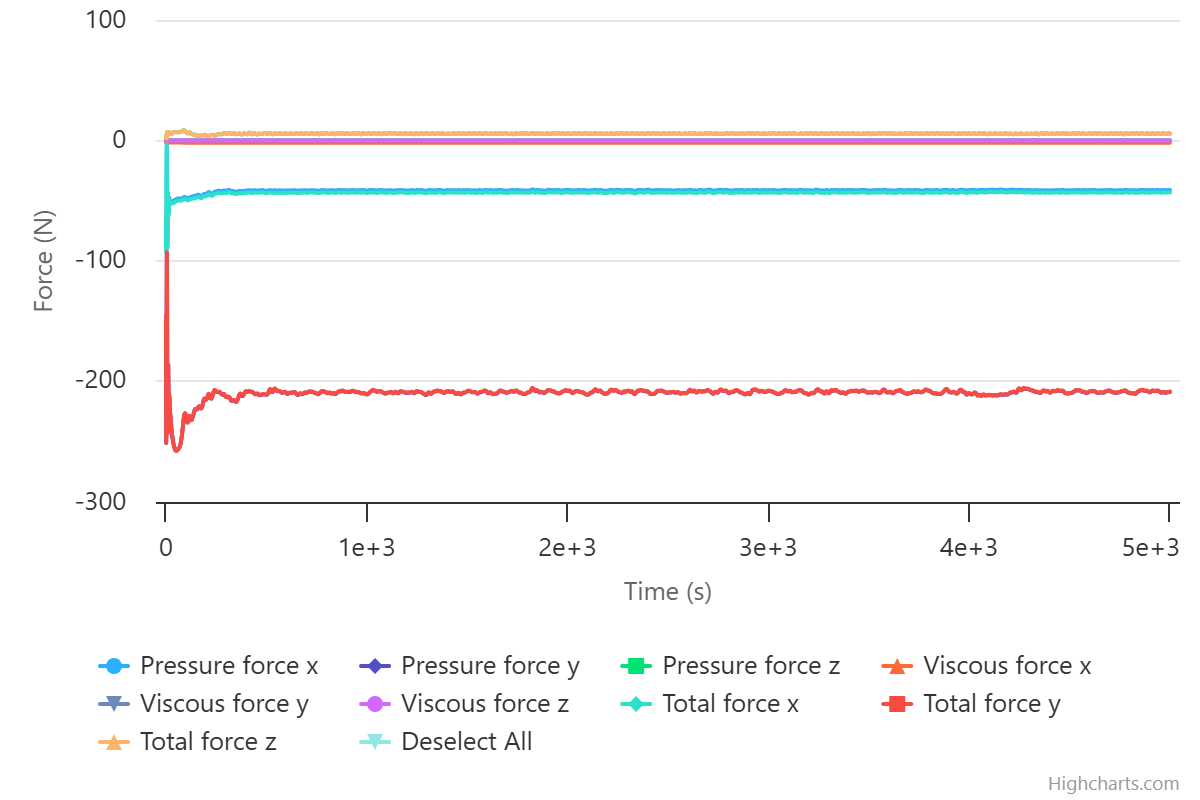}}}
      \caption{Forces and Pressure for \textit{Simulation 1}}
      \label{fig:forces_pressure}
\end{figure}

\section*{APPENDIX}

All data and code are publicly available at \href{https://github.com/NavalShah/F1-Front-Wing-Physics-Informed-NN}{https://github.com/NavalShah/F1-Front-Wing-Physics-Informed-NN}

\section*{ACKNOWLEDGMENT}

I would like to thank rnsomnath on simscale for his model of the F1 Front Wing. I was able to use his model to run simulations and extract relevant data for training and testing.

\bibliographystyle{plain}
\bibliography{references}

\end{document}